# Characterization of a laser filament-induced plasma in air at 10 kHz using optical emission spectroscopy

**Malte C. Schroeder[1,*], Robin Löscher[1], Nikita Bibinov[2], Ihor Korolov[2], Peter Awakowicz[2], Thomas Mussenbrock[2], and Clara J. Saraceno[1,3]**

[1]*Chair of Photonics and Ultrafast Laser Science, Ruhr-University Bochum, Universitätsstr. 150, 44801 Bochum, Germany*

[2]*Chair of Applied Electrodynamics and Plasma Technology, Ruhr-University Bochum, Universitätsstr. 150, 44801 Bochum, Germany.*

[3]*Research Center Chemical Science and Sustainability, University Alliance Ruhr, Universitätsstr. 150, 44801 Bochum, Germany.*

**malte.schroeder-w9x@ruhr-uni-bochum.de*

**Abstract:** The increasing availability of high-power Yb-based ultrafast laser-amplifier systems has opened the possibility of air filamentation at high repetition rates >1 kHz. In this new regime, accumulation effects cannot be ruled out, therefore, characterizing the plasma parameters and afterglow plasma-chemical kinetics becomes increasingly relevant. In this work, we use optical emission spectroscopy to measure nanosecond dynamics of gas temperature and electron temperature, species-specific decay times, and electron density of an atmospheric air laser filament produced by high average power femtosecond laser at a high repetition rate of 10 kHz. The molecular excitation mechanisms behind the nitrogen photoemissions are derived from vibrational distributions and temporal behavior of the studied emission bands. The presented diagnostic technique offers a complementary but more holistic measurement approach to optical probe schemes to characterize the laser-filament-induced plasma wake for high repetition rate filaments.

## 1. Introduction

Filamentation of intense ultrafast laser pulses in media occurs when the peak power of the propagating pulse exceeds a critical threshold. Beyond this threshold, the self-focusing of the laser pulse due to the Kerr effect increases to the point where the beam collapses and reaches intensities that are sufficient to drive multi-photon or tunnel ionization. This results in the formation of a transient plasma that evolves on multiple time scales: first with a high electron density and elevated electron temperature [1–3], followed by a slow plasma wake recombination and complex photochemistry on ns-ms time scales that is not accessible to the fs pulse [4–6]. The first fast plasma formation will, in turn, cause a de-focusing of the incident laser pulse, which leads to an interplay between repeated intensity-driven self-focusing and plasma-based de-focusing. This results in an extended transient plasma channel, which recombines within tens of nanoseconds [7–9]. Since the filament and consequent plasma channel are self-propagating and self-healing structures, they are well-suited tools for different applications, such as remote sensing of chemical/biological agents, triggering of water condensation, cloud clearing, and lighting control [10–12]. However, so far, mostly the fast transient behavior of the plasma has been meaningful to the filamentation community, except for a few works [2,13–16], due to the time mismatch between the femtosecond pulse and the rich photochemistry happening at the plasma wake.

However, as the available repetition rates of ultrafast laser systems have seen extremely fast-paced progress in the last decade [17–21], the need for a more accurate picture and experimental characterization of these photochemical processes happening in the plasma after the femtosecond excitation pulse and the resulting change in the plasma characteristics becomes increasingly relevant. When generating filament-driven plasmas at repetition rates exceeding several kHz, or given sufficiently large pulse energies, or a combination of both as in the presented study, the intra-pulse intervals are too short to allow for the full recovery of the propagation medium from the heat deposited during the exothermic plasma recombination process [22–24]. Given sufficient heating, the slow heat diffusion allows for cumulative effects, which can impact the propagation of the pulse, filament dynamics, and plasma generation itself [2,6,9,23,25,26]. Given enough energy or high enough repetition rates, these cumulative heating effects can also induce turbulence when working under atmospheric conditions. These turbulences negatively affect usual optical probing schemes, making several established diagnostic methods (for example interferometry) challenging or unviable. Going further, if the intra-pulse intervals become sufficiently short, the plasma characteristics, like electron temperature, electron density, ratios of molecular species in the plasma, decay, etc., can be affected by the consecutive pulses through direct laser-plasma interactions [5]. Whereas physical effects related to accumulations are known to occur and have been studied [2,25] and even used as a tool to tailor the fs laser-plasma interaction using pre-pulses [5,8,27], the use of many pulses or full pulse trains at high repetition rate remains unexplored – due to the very high average powers needed to fulfill these conditions, which are only becoming available now.

The laser-induced plasma's transient nature on many different timescales and the corresponding complex excitation pathways within it make the experimental characterization of the conditions inside the plasma challenging. So far, the main diagnostic methods applied for the characterization of filament-induced plasma channels were based on current detection [25], fluorescence or absorption detection via photodetectors [28,29], interferometry [30–32], pump-probe spectroscopy [33], etc., which all have advantages and drawbacks. Both the interferometry and fluorescence measurements yield information about the electron density in the plasma. They have been performed at variable excitation conditions but have neglected to give insight into the molecular composition and excitation pathways of the plasma. Pump-probe spectroscopy provides information about the molecular species present in the plasma, but not about the plasma conditions, such as the electron temperature or density. To obtain a full picture of the plasma would require combining several diagnostic methods and running them simultaneously, an approach not yet employed for the investigation of plasmas generated during laser filamentation. Thus, at the time of this publication, reliable excitation models and accurate descriptions of the plasma conditions in transient femtosecond plasma filaments are still absent.

In the presented study, a laser filament-driven plasma is generated at a repetition rate of 10 kHz by focusing femtosecond laser pulses in air. The process can be split in two phases. Primary phase: During the laser pulse, a plasma channel forms through multiphoton ionization. In this phase, free electrons are strongly accelerated via inverse Bremsstrahlung processes, constituting the primary plasma. Secondary phase: After the laser pulse, the dynamics shift. The high-energy electrons, accelerated during the primary phase, rapidly disperse, leaving behind a region with a net positive charge due to ionized species. This charge separation establishes a strong localized electric field. This field is responsible for additional ionization and the formation of a secondary plasma. This post-pulse plasma is characterized by distinct electron density and temperature profiles, differing from the initial plasma channel, and is the focus of our investigation.

The plasma parameters of the secondary plasma are then investigated using an echelle spectrometer and an intensified charge-coupled device (ICCD) camera with a time resolution of 500 ps. A state-of-the-art femtosecond high-average power laser system allows us to generate different filament conditions at this high repetition rate. Based on the molecular nitrogen photoemission's temporal behavior and the emission spectrum's vibrational distribution, an excitation mechanism for nitrogen photoemission is suggested. The gas temperature, electron density, and electron temperature are determined in the frame of a collisional-radiative model accounting for the 100 μs recovery time between laser pulses. These parameters are determined under the assumption of homogeneous distribution in reproducible plasma by averaging over several laser shots. These parameters were determined for different time delays, but without pulse-to-pulse temporal and spatial resolution, as so-called “effective” characteristics [34], which show good overlap with previously recorded plasma filament parameters. The measured parameters can be used to characterize the plasma conditions and afterglow kinetics quantitatively and to determine the gas mixture composition inside the laser-plasma filament, demonstrating that optical emission spectroscopy (OES) represents a holistic approach to the characterization of laser filament plasmas.

## 2. Experimental setup and applied methodology

### 2.1. Experimental setup

The experimental setup is shown in Fig. 1. The pump laser used in our study is a commercial ytterbium-doped thin-disk regenerative amplifier system (*TRUMPF Scientific DIRA 500-10*) with a central wavelength of 1030 nm and generating a maximum average power of 500 W. The system emits pulses at a repetition rate of 10 kHz with pulse durations of approximately 750 fs and a $1/e^2$ beam diameter close to ~10 mm. The maximum available pulse energy is 50 mJ, corresponding to a maximum peak power of approximately 67 GW. During our investigations, the pulse energies were limited using a thin-film polarizer-based attenuation scheme to 2.5 mJ, 5 mJ, and 10 mJ. This was necessary to avoid more pronounced hydrodynamic effects caused by the cumulative air heating by the plasma [25]. An active beam stabilization was implemented before generating the filament to reduce beam pointing instabilities from the source. To accelerate the filamentation process, the laser pulses were focused using a lens with a focus length of $f = 1000$ mm. The lens was installed on a motorized translation stage to allow the positioning of the filament along the propagation axis. This way, the fluorescence maximum could be shifted in front of the ICCD camera and echelle spectrometer fiber entrance to improve the data acquisition. The temporal behavior of the laser-plasma filament is studied using an ICCD camera (*4 Picos, Stanford Computer Optics*) in combination with two band-pass filters.

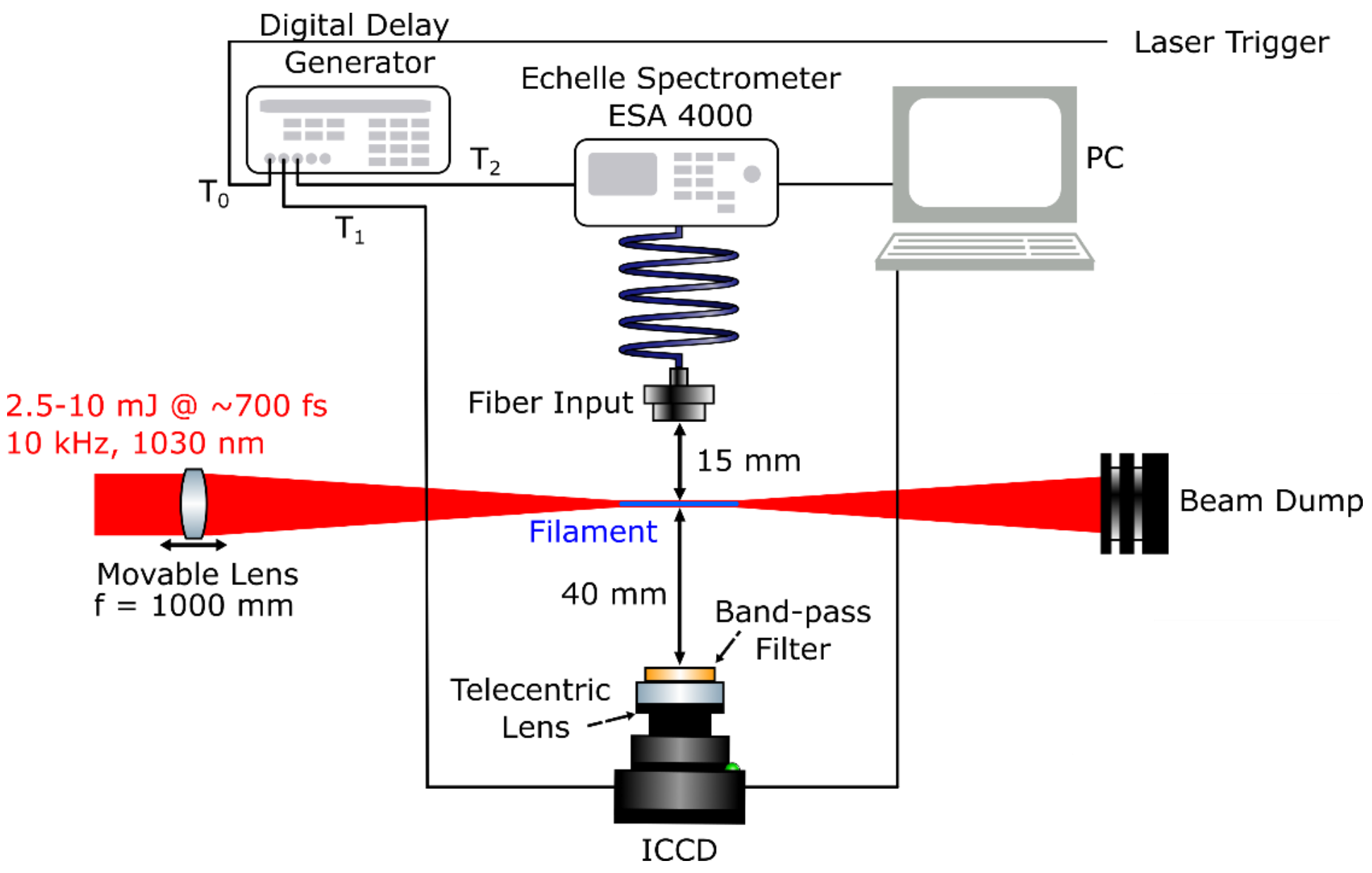

**Fig. 1.** Scheme of the experimental setup for the spectroscopic characterization of the femtosecond laser-plasma filament. Filamentation is accelerated by the use of a focusing lens. The length of the generated plasma filament exceeds the "field-of-view" of both the ICCD camera and the echelle spectrometer.

The exposure time of the ICCD camera can be varied from 200 ps to 1000s with a minimum gate time of 500 ps. Laser-plasma conditions are characterized using nitrogen photoemission $N_2$(C-B) and $N_2^+$(B-X). The two band-pass filters are used with the ICCD.

Their central transmission wavelength and spectral range, (380 ± 5) nm and (390 ± 5) nm are chosen to minimize the difference between the measured signals for the given plasma conditions. To synchronize the camera timing (T1) with the trigger pulse of the laser (T0) and to scan the temporal development of the molecular emission, a digital delay generator (*DG645, Stanford Research Systems*) is interposed between the two.

An echelle spectrometer (*ESA4000, LLA Instruments*) with a spectral resolution of 0.015 nm < $\Delta\lambda$ < 0.06 nm in the spectral range of 200 nm < $\lambda$ < 800 nm is used for the OES-based plasma characterization. The spectral efficiency of the echelle spectrometer was determined using both a tungsten ribbon lamp and a deuterium lamp, each calibrated by the Physikalisch-Technische Bundesanstalt (*PTB, Berlin*) [35]. With the echelle spectrometer, temporally and spatially averaged emission spectra are measured (see Fig. 2). From these spectra, the average gas temperature in the plasma can be determined by using the rotational distribution found in the emission spectrum of the $N_2$(C-B,0-0) or $N_2^+$(B-X,0-0) vibrational bands, at $\lambda = 337.1$ nm and $\lambda = 391.4$ nm, respectively, assuming that the translational and rotational degrees of freedom in the ground state of the nitrogen molecules $N_2$(X) are in equilibrium. The photo-absorption and electron impact excitation of diatomic molecules is limited by the selection rule $\Delta J = 0, \pm 1$. Therefore, the rotational distribution in the excited molecular states is approximately equal to the rotational distribution in the ground state of the molecule. With in-house developed software, the emission spectra of the $N_2$(C-B,0–0) and $N_2^+$(B-X,0-0) vibrational bands are simulated with different rotational temperatures and a spectral resolution equivalent to that of the employed spectrometer. The mean intensities in specific spectral areas of the simulated spectra are determined for different rotational temperatures, normalized to the respective maximum values at $\lambda = 337.1$ nm or $\lambda = 391.4$ nm, and fitted with a polynomial fit. The same evaluation procedure is applied to the measured spectra and a relative mean intensity is determined in the same spectral intervals. The rotational temperature, which in our case is equal to the gas temperature, is evaluated using the relative mean intensity values of the measured spectra and the polynomial fitted to the calculated data. The mean value of the gas temperature and the standard deviation are determined by using several specific spectral intervals in the measured emission spectra. The standard deviation of the mean value is used for the determination of a confidence interval for the measured gas temperature.

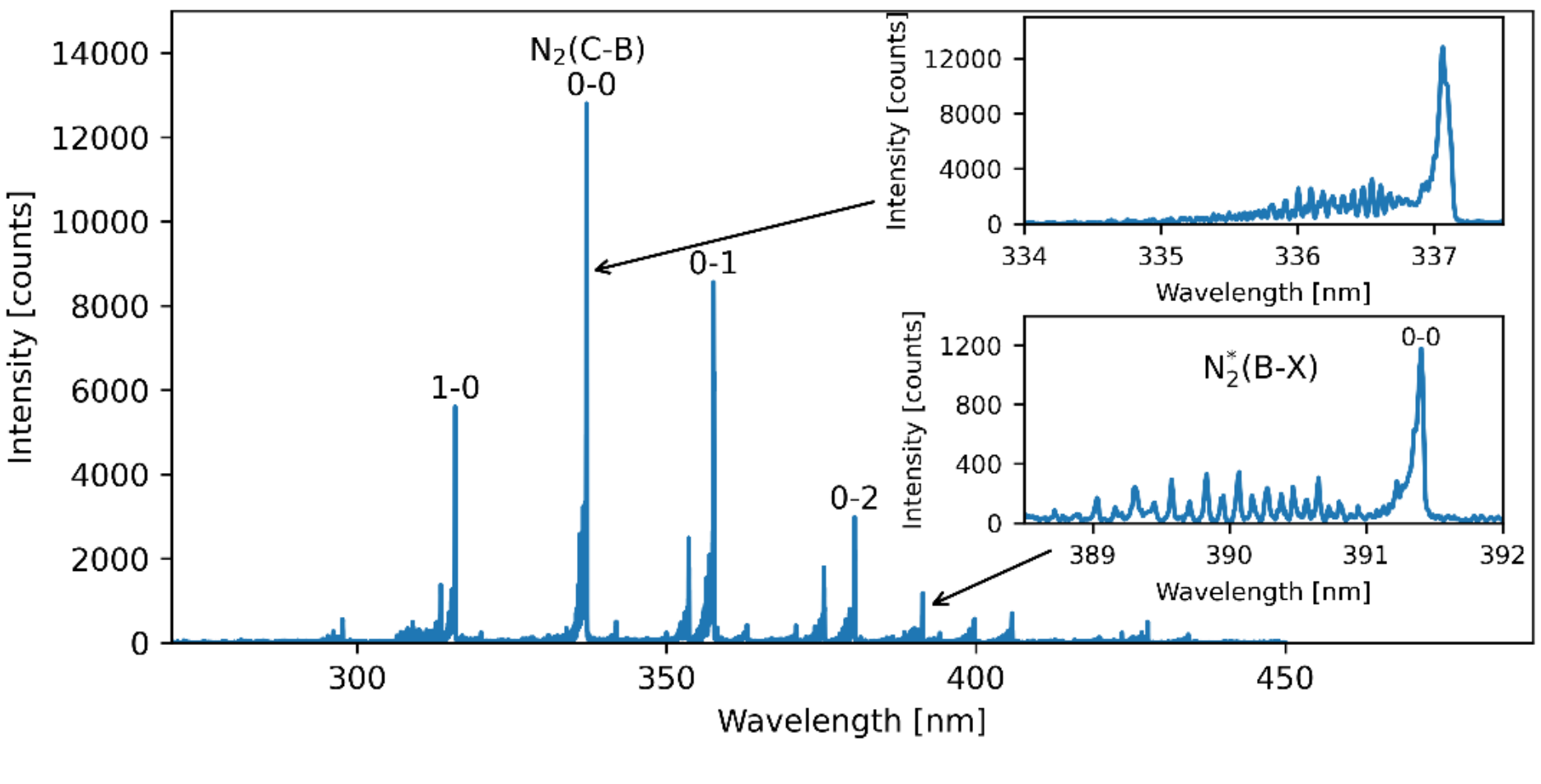

**Fig. 2.** Emission spectrum of the femtosecond laser-plasma filament generated by a 10 mJ pulse. The shown spectrum was measured with the optical arrangement presented in Fig. 1. Observable electronic transitions in molecular nitrogen and the most intensive vibrational bands are presented as insets.

## *2.2. Applied diagnostics method*

In the case of photoemission from molecular nitrogen due to electron impact excitation, the plasma parameters of the laser filament-induced plasma, e.g., electron temperature and electron density, can be determined in the frame of a collisional-radiative model. The emission spectrum of nitrogen under atmospheric pressure conditions in an air plasma contains two intensive emission lines corresponding to the transitions $N_2$(C-B) and $N_2^+$(B–X) (see Fig. 2), which are excited by the impact of electrons with molecular nitrogen in the ground state $N_2$(X). Under this assumption, the electron temperature $k_B T_e$[eV], with $k_B$ being the Boltzmann constant, can be determined based on the measured ratio $R$ of the emission intensities when corrected for collisional quenching:

$$R = \frac{I_{N_2^+(B-X,0-0)} \cdot Q_{CB}}{I_{N_2(C-B,0-0)} \cdot Q_{BX}} = \frac{n_{N_2} n_e k_{N_2^+(B-X)}}{n_{N_2} n_e k_{N_2(C-B)}} = \frac{k_{N_2^+(B-X)}}{k_{N_2(C-B)}} \tag{1}$$

where $Q_{CB}$ and $Q_{BX}$ are the quenching factors for the respective excited states under plasma conditions, $n_{N_2}$[cm$^{-3}$] is the density of molecular nitrogen, $n_e$[cm$^{-3}$] is the electron density, and $k_{N_2^+(B-X)}$ and $k_{N_2(C-B)}$ [cm$^3$s$^{-1}$] the rate constants of the respective electron impact excitations (see Fig. 3). The rate constants are calculated under the assumption that they adhere to a Maxwellian electron energy distribution function (EEDF)

$$k = 2\sqrt{\frac{2C}{\pi m_e}} (k_B T_e)^{-3/2} \int_0^\infty exp\left(-\frac{E}{k_B T_e}\right) E \sigma_{exc}(E) dE, \tag{2}$$

where C=1.6·10$^{-12}$ g·cm$^2$s$^{-1}$eV$^{-1}$, $m_e$[g] is the electron mass, and $\sigma_{exc}$(E)[cm$^2$] is the cross-section of the electron impact excitation given by Itikawa [36]. The electron density is determined using the measured intensity $I_{N_2(C-B,0-0)}[\frac{\text{photon}}{\text{s·cm}^3}]$ of the $N_2$(C-B,0-0) transition and the excitation rate constant $k_{N_2(C-B)}$.

$$n_e = \frac{I_{N_2(C-B,0-0)}}{n_{n_2} \cdot k_{N_2(C-B)} \cdot Q_{CB}} \tag{3}$$

The quenching factors $Q_{CB}$ and $Q_{BX}$ are calculated using Eqs. (4) and (5) via the Einstein coefficients $A_{N_2(C)}$ and $A_{N_2^+(B)}$ of their respective emission transitions, $N_2$(C-B) and $N_2^+$(B-X), and the rate constant of the collisional quenching by molecular nitrogen and oxygen, $k_{qN_2}^{N_2(C)}, k_{qN_2}^{N_2^+(B)}$ and $k_{qO_2}^{N_2(C)}, k_{qO_2}^{N_2^+(B)}$ from Valk et al. [37]. The influence of the elevated gas temperature on the efficiency of the quenching process is accounted for by the substitution of the rate constants for the collisional-induced deactivation in the Arrhenius form.

$$Q_{CB} = \frac{A_{N_2(C)}}{A_{N_2(C)} + k_{qN_2}^{N_2(C)} \cdot \sqrt{\frac{Tg}{300K}} \cdot n_{N_2} + k_{qO_2}^{N_2(C)} \cdot \sqrt{\frac{Tg}{300K}} \cdot n_{O_2}} \tag{4}$$

$$Q_{BX} = \frac{A_{N_2^+(B)}}{A_{N_2^+(B)} + k_{qN_2}^{N_2^+(B)} \cdot \sqrt{\frac{Tg}{300K}} \cdot n_{N_2} + k_{qO_2}^{N_2^+(B)} \cdot \sqrt{\frac{Tg}{300K}} \cdot n_{O_2}} \tag{5}$$

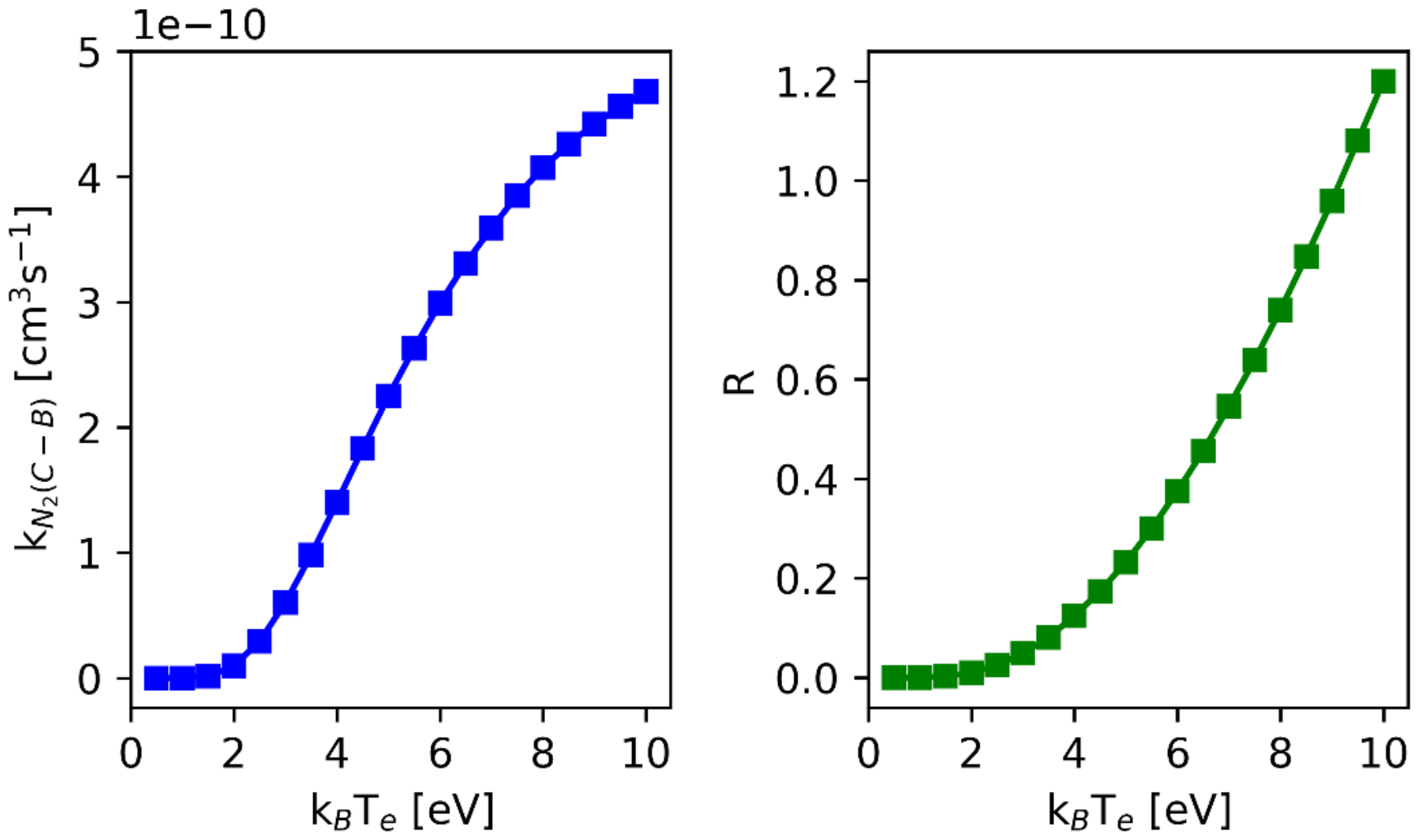

**Fig. 3.** Rate constant for electron impact excitation of $N_2$(C-B,0-0) emission band (left) and ratio R of the emission intensities corrected for the collisional quenching (1) (right) versus electron temperature ($k_BT_e$ in eV).

## 3. Experimental results

### *3.1. Excitation model of nitrogen photoemission*

Plasma formation during femtosecond laser filamentation involves not only direct multiphoton ionization, but also subsequent electron acceleration and secondary ionization processes that lead to complex plasma dynamics [1,4,38–40]. These additional mechanisms highlight the intricate interplay between the laser field and plasma evolution, extending beyond the timescale and simplicity of initial ionization events. For a quantitative determination of the gas composition, the plasma parameters, such as electron density and electron temperature, and gas temperature of the plasma filament, are needed [9,41]. The collisional-radiative model used in this study for the determination of said parameters of the secondary plasma works under the assumption that MPI is the dominant ionization pathway in the primary plasma, based on the calculated Keldysh parameter ($\gamma \approx 2.3$) for our laser parameters and the air medium [42].

The emission spectrum of the plasma filament consists mainly of two molecular nitrogen bands, $N_2$(C-B) and $N_2^+$(B-X) [29,43]. Based on the three-step model presented above, these emission bands can be excited by multiphoton excitation and ionization (step a) or by electron impact excitation of nitrogen molecules (step c) after the electrons are accelerated through the inverse Bremsstrahlung process (step b). The photoemission from the nitrogen molecules and molecular ions can be used to characterize the plasma conditions within the filament (see Section 2.2).

The first question that must be clarified relates to the excitation mechanism behind the nitrogen photoemission from the plasma filament. As shown before, an electronically excited state of the ionized molecular nitrogen can be populated via MPI during the propagation of the laser pulse [29] and additionally through electron impacts in the secondary plasma [44]. The $N_2$(C)-state can be excited by electron-ion recombination [45] in the secondary plasma. A reliable excitation model can be identified by using both the photoemission's temporal behavior and spectral distribution. In an excitation model with multiphoton absorption, the $N_2^+$(B)-state is excited directly during the laser filament propagation (6). In this excitation model, the $N_2$(C)-state is populated with some delay by the formation of $N_4^+$ and electron-ion neutralization (Eqs. (7-9)) [45,46].

$$N_2(X) + n \times h\nu \rightarrow N_2^+(X, A, B, C) + e \quad (6)$$

$$N_2^+(X) + N_2 + N_2 \rightarrow N_4^+ + N_2 \quad (7)$$

with $k_7 = 5.5 \times 10^{-29} \left(\frac{300K}{T_g}\right)^{1.7} cm^6 s^{-1}$ [37]

$$N_4^+ + e \rightarrow N_2(A, B, C) + N_2(X) \quad (8)$$

with $k_8 = 2 \times 10^{-6} \left(\frac{300K}{T_e}\right)^{1/2} cm^3 s^{-1}$ [36]

$$N_4^+ + O_2 \rightarrow O_2^+ + N_2(X) + N_2(X) \quad (9)$$

with $k_9 = 2.5 \times 10^{-10} cm^3 s^{-1}$ [36]

The effective lifetimes of the $N_2(C)$- and $N_2^+(B)$-states ($\tau_C, \tau_B$) (see Eq. (10) and Eq. (11)) in the laser filament-induced plasma are 630 ps and 120 ps, respectively. The lifetime of $N_2^+(X)$ ions under the same plasma conditions is determined by the ion conversion process (Eq. (7)) and amounts to about 200 ps. The lifetime of the $N_4^+$ ions is determined by the charge exchange process (Eq. (9)) and amounts to about 1.2 ns. Based on these values, we can conclude that the emission of the $N_2$(C-B) transition is delayed by about 2 ns when compared to the $N_2^+$(B-X) emission under these plasma conditions.

$$\tau_C = \frac{1}{A_{N_2(C)} + k_{qN_2}^{N_2(C)} \cdot \sqrt{\frac{T_g}{300K}} \cdot n_{N_2} + k_{qO_2}^{N_2(C)} \cdot \sqrt{\frac{T_g}{300K}} \cdot n_{O_2}} \quad (10)$$

$$\tau_B = \frac{1}{A_{N_2^+(B)} + k_{qN_2}^{N_2^+(B)} \cdot \sqrt{\frac{T_g}{300K}} \cdot n_{N_2} + k_{qO_2}^{N_2^+(B)} \cdot \sqrt{\frac{T_g}{300K}} \cdot n_{O_2}} \quad (11)$$

The temporal behavior of the $N_2$(C-B) and $N_2^+$(B-X) emission bands in the filament-induced plasma was measured with the ICCD camera for different laser pulse energies and is presented in Fig. 4. As mentioned before, the temporal resolution of our ICCD camera is 500 ps and, therefore, exceeds the lifetime of the $N_2^+(B)$-states. As such, the temporal behavior of the $N_2^+$(B-X) emission line shows the excitation process while limited by the temporal resolution of the ICCD camera. The first feature, which can be observed in Fig. 4, is the decay constant of the $N_2^+$(B-X) emission, which amounts to approximately 500 ps. This contradicts the suggestion that the photoexcitation of the $N_2^+(B)$-states occurs directly during the femtosecond laser pulse propagation (750 fs) because the effective lifetime of the $N_2^+(B)$-state is considerably shorter – 120 ps. The second observable feature is the delay between the $N_2$(C-B) and $N_2^+$(B-X) emission bands, which appears approximately constant for all pulse energies and amounts to about 500 ps. The delay can be explained by the difference in the effective lifetime of the excited states. We do not observe any additional decay near the 2 ns mark, which can be expected in the frame of the collisional-radiative model with the photoexcitation of the $N_2^+(B)$-state and the production of $N_2(C)$ in electron-ion recombination processes (Eqs. (7-9)). Based on the measured temporal behavior of the nitrogen emission bands, we can assume that both $N_2$(C-B) and $N_2^+$(B-X) emissions are excited in the secondary plasma and not via direct photo-excitation during pulse propagation.

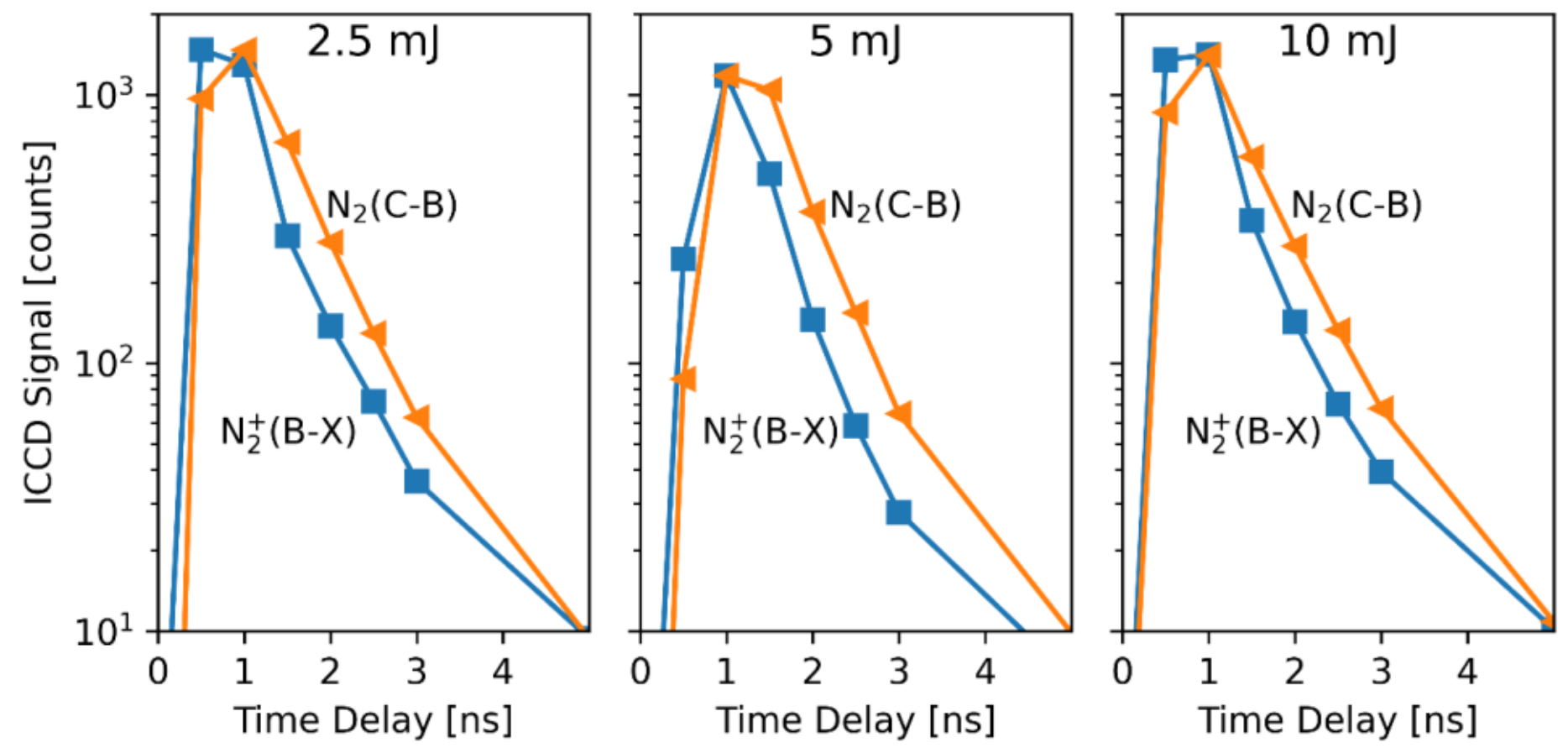

**Fig. 4.** Temporal behavior of normalized $N_2$(C-B) (■) and $N_2^+$(B-X) (◀) fluorescence emission signals measured with an ICCD camera for different laser pulse energies. Time zero of the delay corresponds to the arrival of the laser pulse. A digital delay generator sets the synchronicity between the laser trigger and the camera.

The vibrational excitation of the $N_2$(C)-state visible in the emission spectrum of the filament plasma (see Fig. 2) can help to establish the excitation model. Photo-absorption and electron impact excitation cause vibrational excitations of diatomic molecules according to the Franck-Condon principle, which states that the transition probability scales with the overlap of the vibrational wave functions of upper and lower states. A measure for the overlap is given by the Franck-Condon Factors (FCF). On the contrary, the vibrational distribution of diatomic molecules, which is produced by the dissociation of an excited polyatomic molecule, depends mainly on an energetic balance. The formation of the $N_2$(C)-state by electron-ion recombination (Eq. (6)) was studied by van Koopen et al. [46]. Based on the emission spectrum presented by van Koopen et al. [46] and the Franck-Condon Factors for the $N_2$(C-B) transition from Laux et al. [47], we determine the vibrational population of the $N_2$(C)-state from the emission caused by the electron-ion recombination of $N_4^+$+e (see Fig. 5). This vibrational distribution differs strongly from the vibrational distribution determined using the emission spectra measured from the plasma filament (Figs. 2 and 5). The latter is very similar to the vibrational distribution of the $N_2$(C)-state calculated using the Franck-Condon principle when assuming direct photo-absorption or electron impact excitation from the nitrogen ground state $N_2$(X, v”=0) [48]. Vibrational relaxation, also called V-T or vibrational-translational energy exchange, is negligible under the laser filament-induced plasma conditions because of the sub-nanosecond pulse duration and the significant difference between the gas temperature at about 450 K and vibrational quantum energy of molecular nitrogen at about 3500 K [49]. Excitation via photo absorption from $N_2(X^1\Sigma_g^+)$ to $N_2(C^3\Pi_u)$ is forbidden because of the different spin multiplicity. Based on our data we can thus conclude that the $N_2$(C)-state is excited in the femtosecond laser pulse-induced plasma via electron impact.

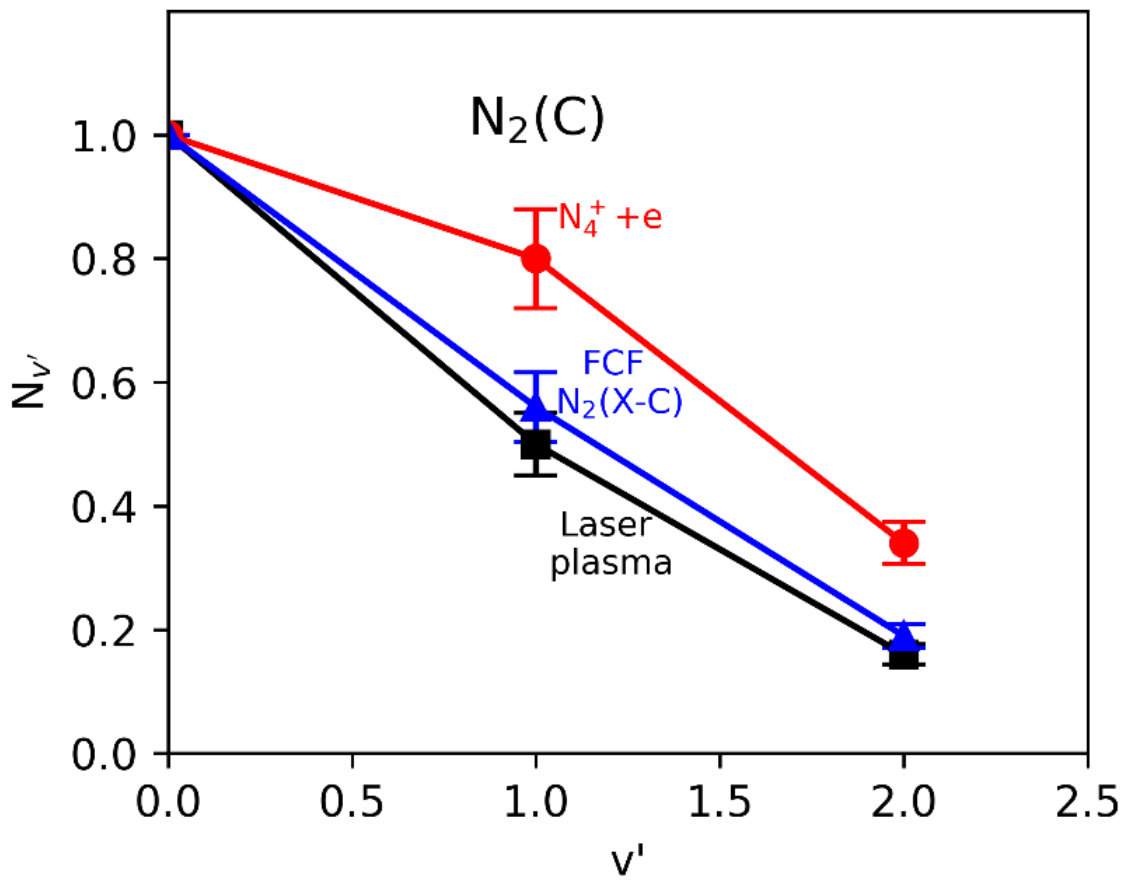

**Fig. 5.** Relative vibrational population of $N_2$(C)-state determined using a spectrum emitted from the electron-ion recombination process $N_4^+$ + e (Eq. (6)) [45] (●, red line), the population measured in the laser-beam plasma (■, black line), and the vibrational population of the $N_2$(C)-state calculated using the Franck-Condon factors of $N_2$(X-C) [48] transition (▲, blue line).

Since the temporal behavior of the nitrogen emission bands and the vibrational distribution in the measured spectra suggest that the $N_2$(C) state in the filament-induced plasma is excited via electron impact, it must occur in the secondary plasma. The measured signal of the $N_2^+$(B-X) emission is delayed relative to the incident laser pulse in our experiment. Thus, we can conclude that excitation for this emission also occurs in the secondary plasma. However, this effect warrants further study under different excitation conditions like different laser repetition rates, average powers, and different photon wavelengths. Further experiments would also benefit from a considerably better temporal resolution.

### *3.2. Laser-plasma characterization*

Under the assumption that both nitrogen emission bands $N_2$(C-B) and $N_2^+$(B–X) are excited in the secondary laser-plasma stage, we can determine the average gas temperature and plasma parameters, like the electron temperature and electron density. Fig. 6 shows the gas temperature in the laser filament-induced plasma for different incident laser pulse energies. The gas temperature was determined using the rotational distribution of the $N_2$(C-B,0-0) and $N_2^+$(B–X,0-0) emission bands (see Fig. 2). Because of the relatively low intensity of the $N_2^+$(B–X) emission band and the significant variance in the measured values the confidence interval of the temperature determination using this emission is significantly larger (see Fig. 6). Based on the data shown in Fig. 6, we can conclude that both emission bands have a similar rotational temperature and that this temperature scales with the incident laser pulse energy. The latter provides an additional argument to the claim that the $N_2$(C) state is excited by electron impact and not by electron-ion recombination. In the case of recombination, a dependence of the rotational distribution of the product of the polyatomic molecule dissociation on the laser pulse energy is very unlikely. The rotational relaxation of the $N_2$(C) state can be excluded at atmospheric pressure conditions in air because of the very effective collisional quenching and very low effective lifetime (Eq. (10)).

The electron temperature in the secondary plasma is determined using the measured emission spectra, Eq. (1), and the data presented in Fig. 3. We follow the assumption that both nitrogen molecular bands $N_2$(C-B) and $N_2^+$(B–X) are excited through electron impact (see Fig. 7). To determine the electron density in the secondary plasma using optical emission spectroscopy (see Eq. (3)), the intensity of the nitrogen photoemission from the $N_2$(C-B) transition must be determined in absolute units. For this purpose, the employed echelle spectrometer needs to be calibrated, and the observed plasma volume must be determined. The acceptance cone of the optical fiber used during the spectroscopy measurements shown in this work is determined in an additional experiment using a point-like light source and a goniometer. A cosine profile is assumed for the acceptance cone of the optical fiber, as the optical fiber collects only a small part of the emitted photons.

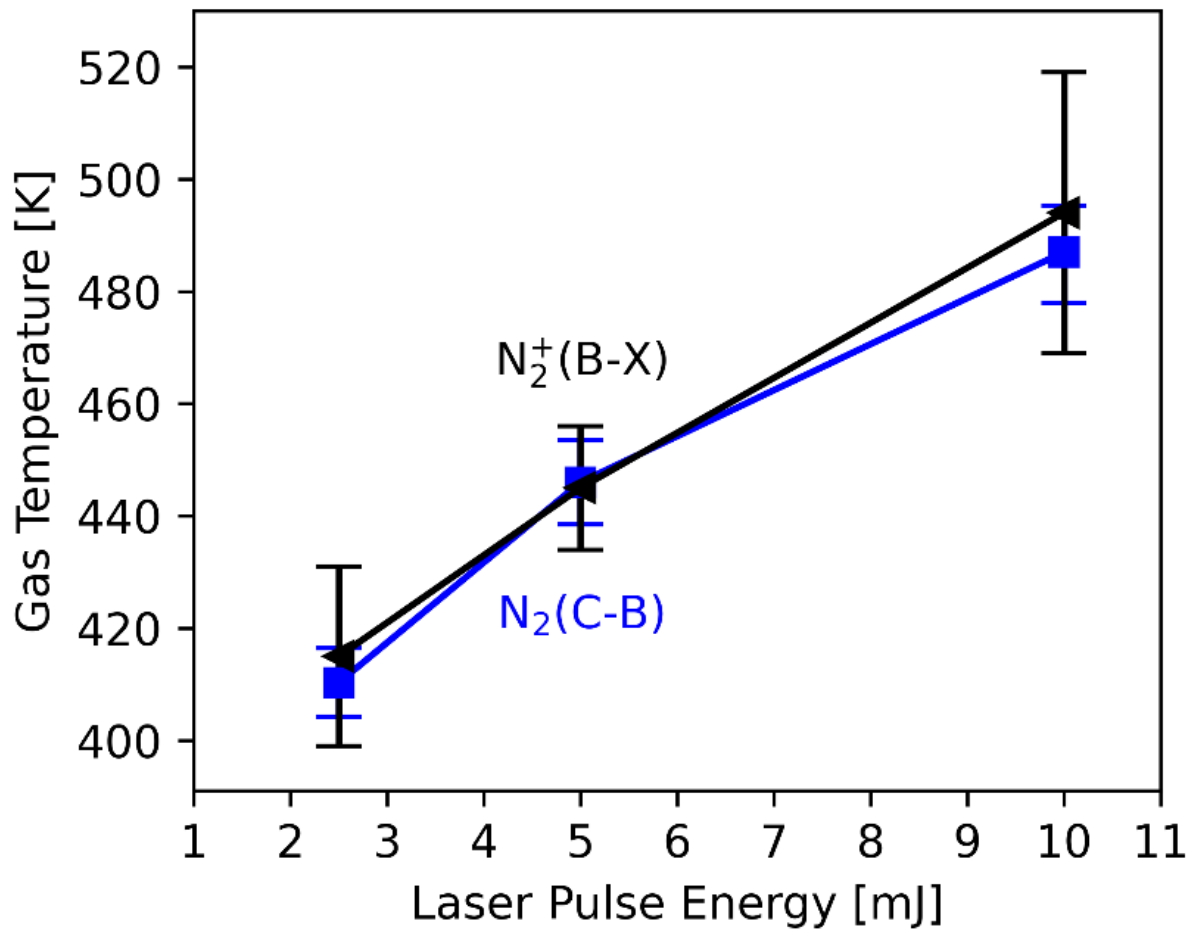

**Fig. 6.** The gas temperature of the laser-plasma filament determined from the rotational distribution measured in the emission spectrum of nitrogen (blue squares-$N_2$(C-B), black triangle-$N_2^+$(B-X)). Confidence intervals of the gas temperature measured using $N_2$(C-B) are considerably lower than that measured with $N_2^+$(B-X) emission.

To take this effect into account during the data evaluation, a geometrical factor $G$ is determined and applied when determining the electron density:

$$G = \frac{s}{4\pi \cdot d^2}, \tag{12}$$

where $s$ is the area of the fiber entrance and $d$ is the distance to the plasma filament. Isotropic photoemission from the secondary plasma of the filament is assumed. Because of the low temporal resolution of the applied spectrometer of 40 ns the ICCD camera is used to determine the fast decay in the photoemission signal (see Fig. 4). Using the FWHM of the measured temporal characteristics, we determine the duration of the plasma emission to be approximately 1 ns.

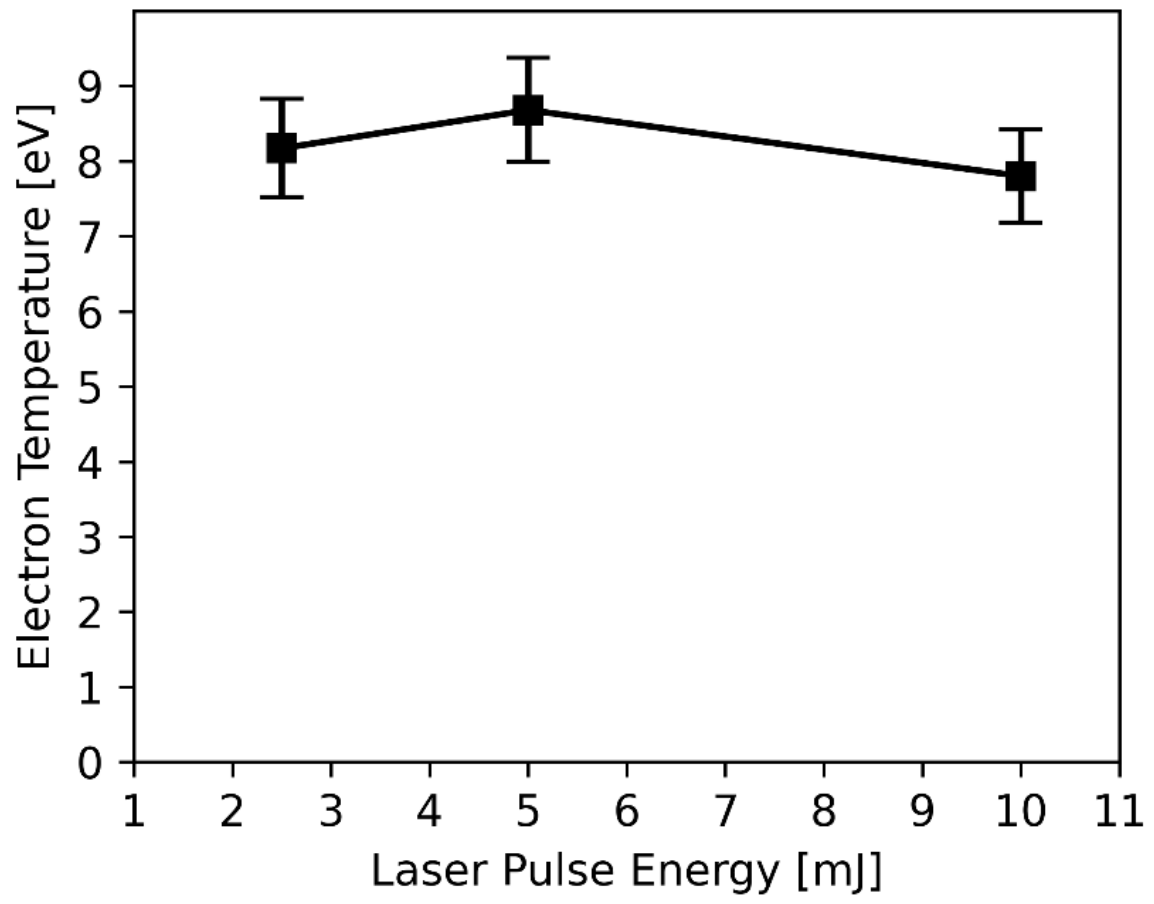

**Fig. 7.** The electron temperature of the laser filament-induced plasma in air determined from the measured nitrogen emission lines and Eq. (1).

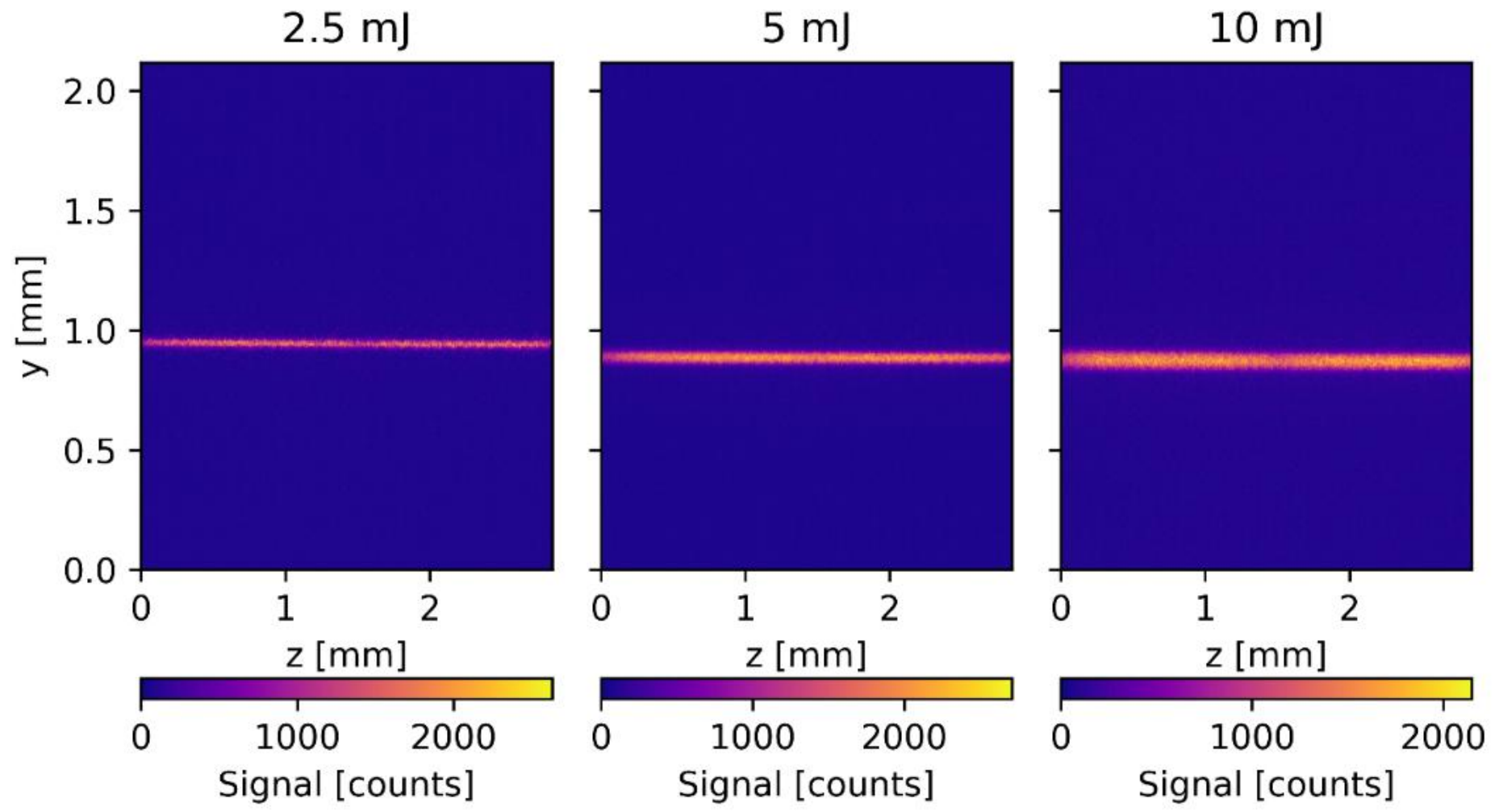

**Fig. 8.** Measured emission intensity frames with 390 nm filter of the laser-induced plasma filament, used to determine the plasma channel profile.

The rate constant for the emission from electron impact excitation of $N_2$(C-B) is calculated using Eq. (2), the cross-section given by Itikawa [36], and the electron temperature values presented in Fig. 7. To determine the density of the molecular nitrogen in the secondary plasma at atmospheric pressure, we use the gas composition of air, the measured gas temperature, and the ideal gas law: $n = \frac{k_B T_g}{p}$. The plasma volume, which is observed with the echelle spectrometer, depends on the length and cross-section of the plasma channel. As mentioned before, the length of the channel is calculated based on the acceptance angle of the optical fiber and the cosine profile of the sensitivity. The plasma channel profile is determined using the ICCD camera for both emission transitions and for all three laser-beam energy values (see Figs. 8 and 9). The measured profiles have a Gaussian-like form (see Fig. 9), following the shape of the intensity profile of the incident femtosecond laser pulses. We use the FWHM of the filament profiles to determine the plasma filament cross-section. The profiles' width for both nitrogen transition emissions depends strongly on the laser pulse energy but are very similar to each other (see Fig. 9). The electron densities determined for different pulse energies from the measured spectra and ICCD images in the collisional-radiative frame are presented in Fig. 10. The accuracy of the presented values is discussed in the next section.

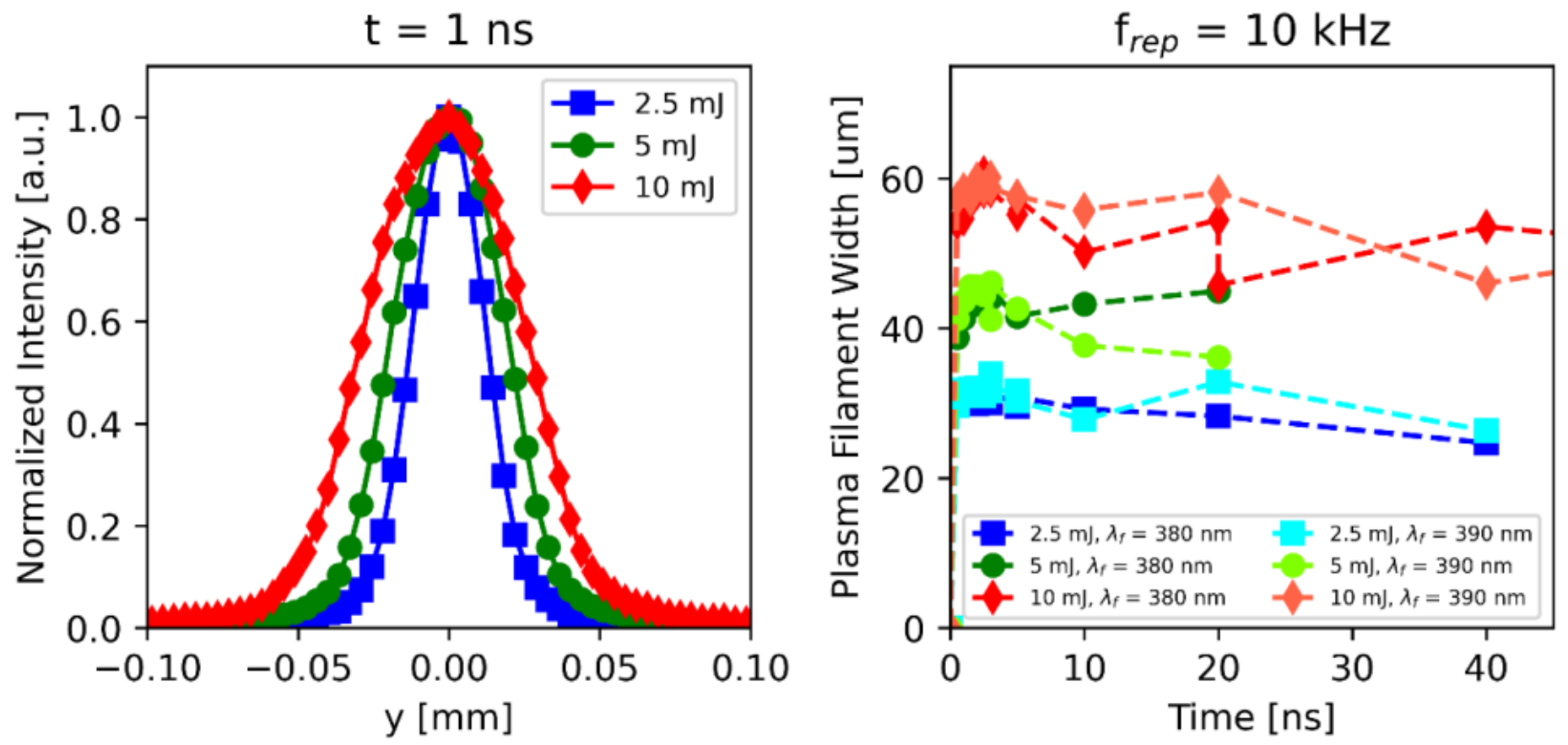

**Fig. 9.** The temporary averaged profiles of the plasma filament measured for the $N_2^+$(B-X) emission (left). Temporally and spectrally resolved values of the FWHM of the measured plasma channel profiles (right).

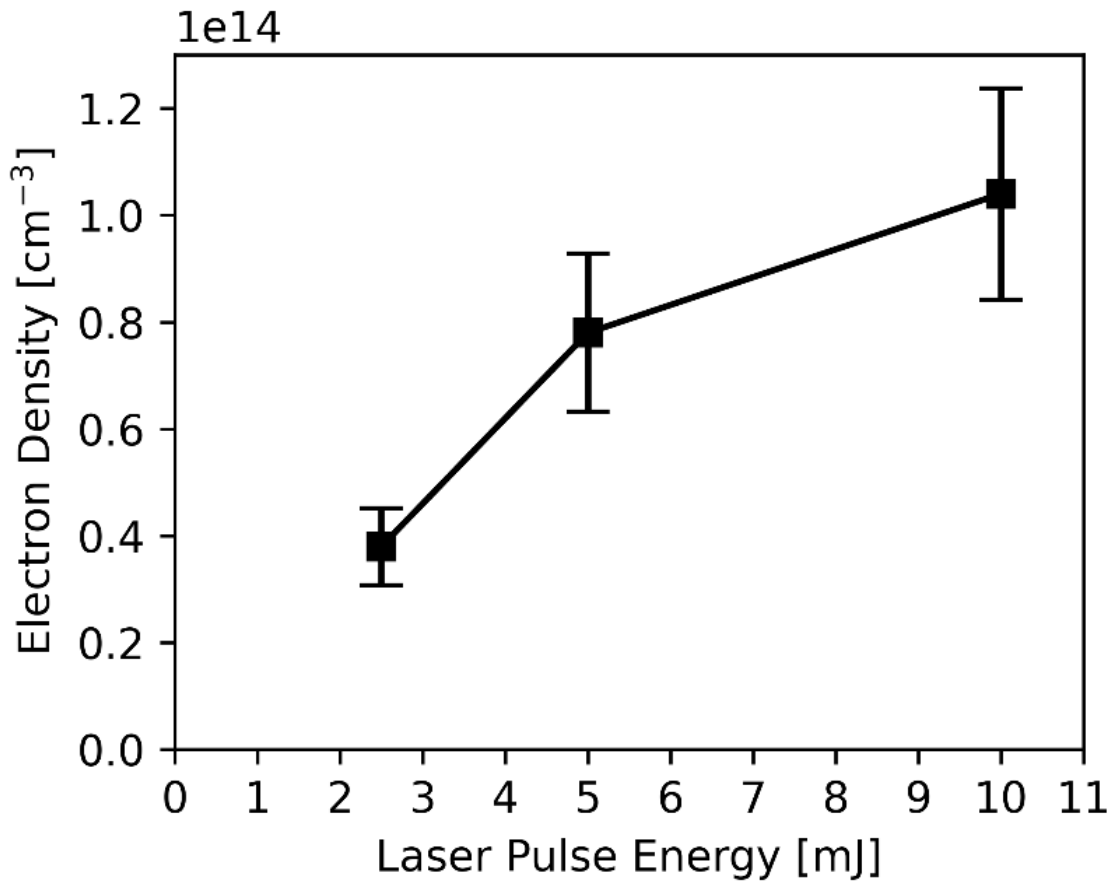

**Fig. 10.** Electron densities in the plasma filament determined from the optical emission spectroscopy data within the framework of the collisional-radiative model. Densities are averaged over the first nanosecond after the incident laser pulse using ICCD estimated duration of the plasma emission.

## 4. Discussion

### *4.1. Excitation of nitrogen photoemission in laser-induced plasma filaments*

In the emission spectrum of the laser-induced plasma filaments, we observe emission bands of molecular nitrogen ($N_2$(C-B) and $N_2^+$(B–X)), which can be excited by multiphoton absorption, electron impact, and electron-ion recombination [1,4,31]. We evaluate the temporal behavior and spectral distribution of the studied emission bands to resolve this ambiguity.

#### *4.1.1. $N_2$(C-B) transition*

A. The multiphoton absorption excitation of nitrogen from $N_2(X^1\Sigma_g^+)$ to $N_2(C^3\Pi_u)$ is forbidden due to the different multiplicity of the lower and upper states. However, electron impact makes this excitation transition possible without violating selection rules.

B. The $N_2(C)$ state cannot be a product of the electron-ion recombination of $N_4^+ + e$ because of the absence of the expected ~1 ns delay in the emission signal caused by the lifetime of the $N_4^+$ ion in air at atmospheric pressure. Additionally, the vibrational distribution measured in the $N_2(C)$ state, which corresponds to an expected distribution from electron impact excitation from the ground state $N_2(X)$ [48], differs strongly from a distribution where the vibrational excitation is driven by electron-ion recombination [45].

C. The rotational distribution of the $N_2(C)$ state depends on the laser pulse energy, which is shown by an increase in the gas temperature and through the scaling of the electron impact excitation $N_2(X) + e \rightarrow N_2(C)$, contradicting the expectation of electron-ion recombination via $N_4^+ + e$.

### *4.1.2. $N_2^+$(B–X) transition*

A. The temporal behavior of the $N_2$(C-B) and $N_2^+$(B–X) emission bands is very similar, with a time delay between emissions of about 500 ps, which is expected because of the longer effective lifetime of the $N_2$(C)-state.

B. The spatial distribution of the $N_2$(C-B) and $N_2^+$(B–X) emissions is similar and equally dependent on the laser pulse energy (see Fig. 9). We do not observe any distinct core inside of the plasma channel, instead, it follows the Gaussian beam profile of the incident laser pulses, which is expected from multiphoton excitation [1].

Based on the above-presented data, we can conclude that both the $N_2$(C-B) and $N_2^+$(B–X) emission bands are excited by electron impact in the secondary plasma. We do not observe any evidence in our analysis that contradicts this statement.

### *4.2. Applied collisional-radiative model*

The gas temperature and plasma parameters of the secondary plasma are determined using optical emission spectroscopy in the frame of a collisional-radiative model. This model assumes steady-state conditions for the Maxwellian electron energy distribution function. On the other hand, multiphoton ionization and the inverse Bremsstrahlung process produce electrons with elevated kinetic energy, which is spent through elastic and inelastic scattering with other electrons and heavy species. The Maxwellian distribution function is a result of elastic collisions of electrons with the surrounding species. The application of the Maxwellian distribution function via the characterization of the transient conditions of the secondary plasma is a rough approximation. However, this characterization can help to understand the physics of the studied processes through comparison with theoretically calculated distribution functions for these conditions, or predicted probabilities of some plasma-chemical processes, which use electrons carrying a kinetic energy higher than 11 eV, which marks the excitation threshold of the $N_2$(C) state. Many atoms and molecules can be excited or dissociated in this energy range. An additional argument to apply to the Maxwellian distribution function via the characterization of the secondary plasma is the probability of elastic collisions, which is considerably higher than the probability of inelastic collisions [36]. This causes a partial Maxwellization of the transient electron distribution function formed by the femtosecond laser excitation before the first inelastic collisions occur.

### *4.3. Accuracy of the plasma parameters determination*

The number of sources of inaccuracy for the determination of the plasma parameters is large, but fortunately, the impact on the parameters from many of the possible errors is relatively low and can be neglected. The accuracy of the gas temperature determination using the rotational distribution in the emission spectra of the diatomic molecules depends mainly on the total signal level, signal-to-noise ratio, and spectral resolution of the echelle spectrometer. Due to considerably higher intensities in the emission spectrum of the filament plasma, the gas temperature determination using the $N_2$(C-B,0-0) vibrational band (relative error 2-3%) is more precise than when using the $N_2^+$(B-X) emission (error about 10%).

The inaccuracy in the determination of the electron temperature (8%) is derived by using quadratic error propagation of $R(k_B T_e)$ in Fig. 3 (right), the inaccuracy of the relative calibration of the spectrometer in this spectral range(5%) [35], the inaccuracy of the given cross-sections for the electron impact excitation $N_2$(C-B) (14%) and $N_2^+$(B-X) (10%) [36], and the collisional quenching $Q_{CB}$ (2%) and $Q_{BX}$ (10%) [37]. The inaccuracy of the electron density determination (19%) stems from the inaccuracy in the relative calibration of the spectrometer (5%), the absolute calibration of the spectrometer (8%) [35], the electron temperature (8%), the given cross-section of electron impact excitation (14%), the gas temperature (3%), and the quenching factor $Q_{CB}$ (2%).

### *4.4. Comparison with other studies*

In several experimental and theoretical studies, plasma filament conditions were characterized for different laser wavelengths and laser pulse energies (see, e.g. [1,16,30,31,50,51]). The densities of the free electrons ($10^{12}$ cm$^{-3}$ – $10^{18}$ cm$^{-3}$) generated in the laser-induced plasma were determined with time-resolved diffraction measurements, using plasma conductivity measurements, and through theoretical studies. The experimentally retrieved electron density, shown in Figure 10, is lower than the expected values of $10^{15}$ cm$^{-3}$ – $10^{16}$ cm$^{-3}$ from the literature. This is most likely due to two reasons: The laser system used operates at a longer wavelength than the most common ones used in filamentation experiments so far. It emits light at 1030 nm instead of 800 nm, which decreases the ionization cross-section for oxygen and nitrogen, resulting in a lower MPI yield. Secondly, due to the increased repetition rate, the cumulative gas heating effects mentioned in the introduction negatively affect the ionization rate. In the permanently heated filamentation region, the air density is reduced, leaving less ionizable material. Additionally, the nonlinear refractive index, proportional to the air density, is diminished, resulting in a weaker Kerr response. Overall, this leads to a reduction in electron density.

Filament plasma diameters (FWHM) found in the literature varied from 20 µm – 100 µm, as they depend on a multitude of parameters, such as pulse duration, pulse energy, focus geometry, propagation medium, etc. The values we retrieved from the ICCD measurements fall in this range, as shown in Figure 9. The electron temperature of the filament plasma was previously estimated to be approximately 5000-5800 K (~0.5 eV). It is worth noting that these values were mostly measured for filament plasmas generated from sub-50 fs pulses. For short pulses, inverse Bremsstrahlung has only a minor effect on the electron energy [52]. For longer pulses above 100 fs, such as the ones emitted by our laser systems, electrons can acquire significant kinetic energy from the inverse Bremsstrahlung, enough to allow impact ionization or avalanche ionization of surrounding molecules and atoms [44]. The determined electron temperatures of about 8 eV, shown in Figure 7, can thus be explained by this effect. The determined gas temperatures of roughly 410 K to 450 K and 490 K match previously measured and simulated temperatures by Cheng et al. [22].

Optical emission spectroscopy is employed in the presented study to characterize the plasma conditions. Only two emission bands, $N_2$(C-B) and $N_2^+$(B–X), are used in this method, and therefore, a profile for the electron distribution function must be assumed initially or additionally determined through simulations. The Maxwellian distribution function is assumed for this purpose for a low-pressure plasma [53], and the electron destruction function is simulated by solving the Boltzmann equation at elevated pressure conditions [35]. The latter method cannot be used to characterize the filament plasma conditions because of non-steady-state conditions and the fact that electrons can be partially accelerated in the primary plasma by the inverse Bremsstrahlung process. As mentioned previously, we assume that free electrons are accelerated during laser pulse excitation, and a non-steady-state electron distribution function is formed. Because of the very high cross-section for elastic collisions between electrons and between electrons and heavy particles, Maxwellization of the electron distribution function occurs before other physical processes, such as excitation, dissociation, and ionization of diatomic molecules through inelastic collisions. This assumption must be verified by comparison of the experimental results with theoretical simulations.

The plasma parameters, which are determined using the $N_2$(C-B) and $N_2^+$(B–X) emission lines, characterize the electron energy distribution function in the kinetic energy range above the threshold of the $N_2$(C-B) excitation at 11 eV. The extrapolation of the electron energy distribution function in the energy range below this threshold is unreliable and can cause a systematical error when determining the plasma parameters, like the electron temperature and electron density of cold electrons E< 10 eV. However, the plasma parameters determined using the nitrogen photoemission lines can be used to reliably calculate the rate of the plasma chemical reactions caused by impacts of electrons with kinetic energies higher than 11 eV, such as excitation, dissociation, and ionization.

## Conclusion

This work represents a foray into determining the relevant parameters and photochemical processes occurring in a plasma generated during the filamentation of an ultrafast laser pulse in air. Particular emphasis is placed on understanding the formation and characteristics of the post-pulse secondary plasma, which is critical for applications requiring precise control of plasma properties, such as in materials processing, spectroscopy, and advanced photonic device development. The gas temperature and plasma parameters of the plasma, generated from the focusing of 750 fs laser pulses at a repetition rate of 10 kHz and a central wavelength of 1030 nm in air, are determined using molecular nitrogen photoemission lines in the framework of a collisional-radiative model without temporal and spatial resolution. The laser pulse energy was varied from 2.5 mJ to 5 mJ and 10 mJ during this study. The diameter of the plasma filament (FWHM) spans from 30 μm to 45 μm and 60 μm depending on the laser pulse energy. The gas temperature increases with the laser pulse energy from 410 K to 450 K and 490 K. The electron temperature is approximately constant at 8 eV, while the average electron density increases with the laser pulse energy from $4 \cdot 10^{13}$ cm$^{-3}$ to $8 \cdot 10^{13}$ cm$^{-3}$ and $10^{14}$ cm$^{-3}$. The influence of non-steady-state conditions on the reliability of the plasma parameters is discussed and compared to established diagnostic methods.

## Outlook

In future work, as established by Magazova et al. [41], a complete characterization of a plasma filament must include a full 3D characterization. Optical emission spectroscopy in the frame of a collisional-radiative model in combination with the Abel inversion of the images measured with a CCD camera can help to solve this problem [35,53]. Additionally, future studies would greatly benefit from theoretical simulations of the formation of the electron distribution function in the inverse Bremsstrahlung process and its temporal evolution up to the occurrence of photon emission.

In terms of experiments, we propose a full parametric study of the plasma parameters and excitation processes using OES in the frame of a collisional-radiative model for different laser pulse and experimental parameters. Conclusive studies on the nature of the filament plasma are still lacking and tend to be disjointed, focusing on or giving insight into only singular aspects of the plasma or plasma generation process, with a few notable exceptions [2,13-16]. Given the holistic nature of the OES approach, a more complete picture of plasma filament generation could be achieved by repeating the measurements presented in this work for more pulse energies, pulse durations, focus geometries, laser repetition rates, wavelengths, and gas mixtures. Ultimately, investigating changes to the emission spectra when reaching laser pulse repetition rates that approach the lifetimes of excited states in the plasma will be essential as more substantial chemical changes are expected to occur there [5,8,27].

## Acknowledgment

This work was also supported by the German Research Foundation in the frame of the collaborative research center SFB 1316, project A5, under Project-ID 434319380, and under Germany's Excellence Strategy-EXC-2033-Projektnummer 390677874 - RESOLV. We acknowledge support by the DFG Open Access Publication Funds of the Ruhr-Universität Bochum.

## AUTHOR DECLARATIONS

### Conflict of Interest

The authors declare no conflict of interest.

### Data availability

Data underlying the results presented in this paper is publicly available on the Zenodo platform: 10.5281/zenodo.14843545 **[54]**.